# Mechanism of Polarization Fatigue in BiFeO$_3$: the Role of Schottky Barrier

Yang Zhou,[1] Xi Zou,[1] Lu You,[1] Rui Guo,[1] Zhi Shiuh Lim,[1] Lang Chen,[1] Guoliang Yuan,[2,a)] and Junling Wang[1,b)]

[1]*School of Materials Science and Engineering, Nanyang Technological University, Singapore 639798, Singapore*

[2]*Department of Materials Science and Engineering, Nanjing University of Science and Technology, Nanjing 210094, China*

## ABSTRACT

By using piezoelectric force microscopy and scanning Kelvin probe microscopy, we have investigated the domain evolution and space charge distribution in planar BiFeO$_3$ capacitors with different electrodes. It is observed that charge injection at the film/electrode interface leads to domain pinning and polarization fatigue in BiFeO$_3$. Furthermore, the Schottky barrier at the interface is crucial for the charge injection process. Lowering the Schottky barrier by using low work function metals as the electrodes can also improve the fatigue property of the device, similar to what oxide electrodes can achieve.

[a)] Electronic mail: yuanguoliang@mail.njust.edu.cn

[b)] Electronic mail: jlwang@ntu.edu.sg

Ferroelectric fatigue refers to the decrease of switchable polarization in a ferroelectric material after repetitive electrical cycling.[1] It is detrimental to ferroelectric based devices and should be minimized. There have been a large number of reports on the mechanism of the polarization fatigue.[2-5] The models proposed can generally be classified into three categories, namely charge injection,[6-9] defects, i.e. oxygen vacancies, redistribution[10-14] and local phase decomposition.[15-17]

In our previous report, we have demonstrated that charge injection is likely the cause of fatigue in $BiFeO_3$.[18] By using a planar capacitor (inset of Fig. 1 (a)), we have conducted piezoelectric force microscopy (PFM) and scanning Kelvin probe microscopy (SKPM) studies to investigate the domain evolution and space charge activities in $BiFeO_3$ during fatigue measurement. A clear correlation between injected electrons at the electrode/$BiFeO_3$ interface and domain pinning is established as shown in Figs. 1(a)-1(c) (For details, please refer to Ref. 18). However, this is not to say that defects are irrelevant to polarization fatigue. On the contrary, these injected electrons must be trapped in gap states which can be associated with existing defects or created by the high energy injected electrons. We emphasize that it is not the redistribution/accumulation of defects, but rather the charging/discharging of defects, leads to polarization fatigue eventually.

After establishing the correlation between charge injection and fatigue, one naturally wonders what controls the charge injection process. Previous studies have demonstrated that oxide electrodes improve the fatigue performance of ferroelectric devices dramatically,[19] so how is charge injection affected when different electrodes are used? We try to clarify this issue by investing planar $BiFeO_3$ capacitors with different electrodes. It was observed that the Schottky barrier at the electrode/$BiFeO_3$ interface plays a significant role in the charge injection process.

By reducing the barrier height, metal electrodes with low work functions can also improve the fatigue performance of $BiFeO_3$.

All the films and electrodes in this work are deposited by pulsed laser deposition (PLD). For the planar devices, 40 nm $BiFeO_3$ thin films are grown at 650 ℃ under 100 mTorr oxygen partial pressure. During the $BiFeO_3$ growth, the laser frequency is fixed at 5 Hz and energy density at ~1 $J/cm^2$. The electrodes are prepared using standard photolithography process and the channel width is 5 μm. Conventional vertical capacitors have also been prepared to further support our conclusion. For this purpose, bottom $(La_{0.7},Sr_{0.3})MnO_3$ electrodes are grown on (001)-oriented $SrTiO_3$ substrate at 800 ℃ under oxygen partial pressure of 200 mTorr followed by $BiFeO_3$ films (~100 nm) deposited at 675 ℃ and under 50 mTorr oxygen partial pressure. Precision LC Ferroelectric tester (Radiant Technologies) is used to measure the polarization and apply bipolar electrical pulses for the fatigue measurement. A commercial atomic force microscope (AFM, Asylum Research MFP3D) is used to conduct the PFM and SKPM scanning and MikroMasch DPE 14 tips (Pt-coated, 160 kHz, and 5.7 N/m) are used.

To clarify the difference between metal and oxide electrodes, we have prepared and investigated $BiFeO_3$ planar capacitors using $(La_{0.7},Sr_{0.3})MnO_3$ as electrodes. All the parameters for $BiFeO_3$ films deposition are the same as reported in our previous study.[18] Since $(La_{0.7},Sr_{0.3})MnO_3$ requires higher deposition temperature than $BiFeO_3$, the electrodes are prepared (by standard lithography and etching) first followed by $BiFeO_3$ (40 nm) deposition. Macroscopic polarization-electric field measurement reveals no fatigue after $10^{10}$ cycles (data not shown). In the in-plane (IP) PFM images taken after opposite electric field is applied to the film, no domain pinning is observed up to $10^{10}$ cycles (Figs. 1(d) and 1(e)). Furthermore, SKPM image reveals negligible electron injection at the $(La_{0.7},Sr_{0.3})MnO_3/BiFeO_3$ interfaces (Fig. 1(f)), whereas significant domain pining and electron injection are found at the $Pt/BiFeO_3$ interface after the same number of switching cycles (Figs. 1(a)-1(c)). The microscopic observation is consistent with previous reports that oxide electrodes can improve the fatigue performance of ferroelectric materials.

Why is there negligible charge injection occurring at the $(La_{0.7},Sr_{0.3})MnO_3/BiFeO_3$ interface? To answer this question, we have to look for the differences between $Pt/BiFeO_3$ and $(La_{0.7},Sr_{0.3})MnO_3/BiFeO_3$ interfaces. From electronic structure point of view, Pt forms a Schottky barrier with $BiFeO_3$[20] while $(La_{0.7},Sr_{0.3})MnO_3/BiFeO_3$ interface is Ohmic. This can be seen from the *I-V* behavior of these two devices shown in Fig. 2(a). By measuring the current vs. temperature behavior, we can calculate the Schottky barrier at the $Pt/BiFeO_3$ interface to be ~1.1 eV (Fig. 2(b)), while the $(La_{0.7},Sr_{0.3})MnO_3/BiFeO_3$ interface shows no barrier. One possible scenario is that the Schootky barrier helps to retain the injected charges at the interface region while an Ohmic contact allows them to move across the interface freely during fatigue measurement. For the $Pt/BiFeO_3/Pt$ planar capacitor, a large voltage drop arises at the reversely biased Schottky interface and external electric field drives electrons to overcome the barrier and inject into the $BiFeO_3$ film during the fatigue measurement (Fig. 2(c)). Once the electrons are injected, they will be trapped at the interface region. When opposite electric field is applied, the same occurs at the other interface and the previously injected electrons remain intact. Eventually, injected electrons accumulate at both interfaces and lead to fatigue. On the other hand, $(La_{0.7},Sr_{0.3})MnO_3$ forms Ohmic contact with $BiFeO_3$ and a flat band with negligible barrier is expected at the interface (Fig. 2(d)). In this case, external field drops uniformly across the whole film. Even though electron injection may still occur, they will be detrapped under opposite field, leading to negligible accumulation under repetitive cycling.

If the above model is correct and the Schottky barrier at the electrode/film interface helps the charge injection process which eventually leads to polarization fatigue, then low work function metals should improve the fatigue performance of $BiFeO_3$ by reducing the Schottky barrier. We have tested this prediction using Fe which has a work function of ~4.5 eV, comparable to that of $BiFeO_3$ (~4.7 eV). $Fe/BiFeO_3/Fe$ planar capacitors are prepared following the same procedure. Much higher leakage current is observed, reflecting the lower interface barriers as expected. After the device is subjected to electrical switching (100 kV cm$^{-1}$ square pulse at 0.1 ms) for $10^{10}$ cycles, we observe no fatigue in the polarization-electric field (*P-E*) loops (Fig. 3(a)). More importantly, the PFM images indicate a very small amount of domain pinning (Figs. 3(c) and 3(d)) and SKPM image shows negligible injected electrons at the interface (Fig. 3(b)). This observation confirms that the Schottky barrier does play a role in the charge injection process, and low work function metals such as Fe should improve the fatigue performance of $BiFeO_3$.

In order to further support our claim, we have conducted fatigue measurement using conventional vertical capacitors. 100 nm $BiFeO_3$ films are deposited on (001)-oriented single crystal $SrTiO_3$ substrates with 20 nm $(La_{0.7},Sr_{0.3})MnO_3$ as the bottom electrode. The schematic structure of the devices is given in Fig. 4(a). Pt, Fe and LSMO are deposited as the top electrodes. They all have the same size of ~100 μm$^2$. When high work function metal electrode Pt is used, fatigue appears relatively soon. Considerable reduction of switchable polarization is observed after only $10^4$ electrical cycles (Fig. 4(b)). For low work function metal electrode (Fe), on the other hand, no fatigue is observed up to $10^7$ cycles, though further cycling leads to break down of the device due to the larger leakage current associated with the low interface barrier. For $(La_{0.7},Sr_{0.3})MnO_3$, no fatigue occurs up to $10^8$ cycles. We thus conclude that Schottky barrier at the electrode/ferroelectric interface indeed plays a role in polarization fatigue and low work function electrodes, metal or oxide, can improve the fatigue performance. However, why Fe electrode makes the device more prone to breakdown is still under investigation.

In summary, following our previous study, where charge injection at the electrode/ferroelectric interface was identified as the cause of fatigue in $BiFeO_3$, we have conducted further experiments and demonstrated that the Schottky barrier at the interface plays a significant role. By reducing the barrier height, even metal electrodes can significantly improve the fatigue performance of ferroelectric devices. This conclusion is corroborated by investigation using conventional vertical capacitors, and is valuable for better understanding and application of ferroelectric materials.

**Acknowledgements:**

This work is supported by National Research Foundation of Singapore under project NRF-CRP5-2009-04. G.L.Y. acknowledges support from National Natural Science Foundation of China under project no. 11134004.

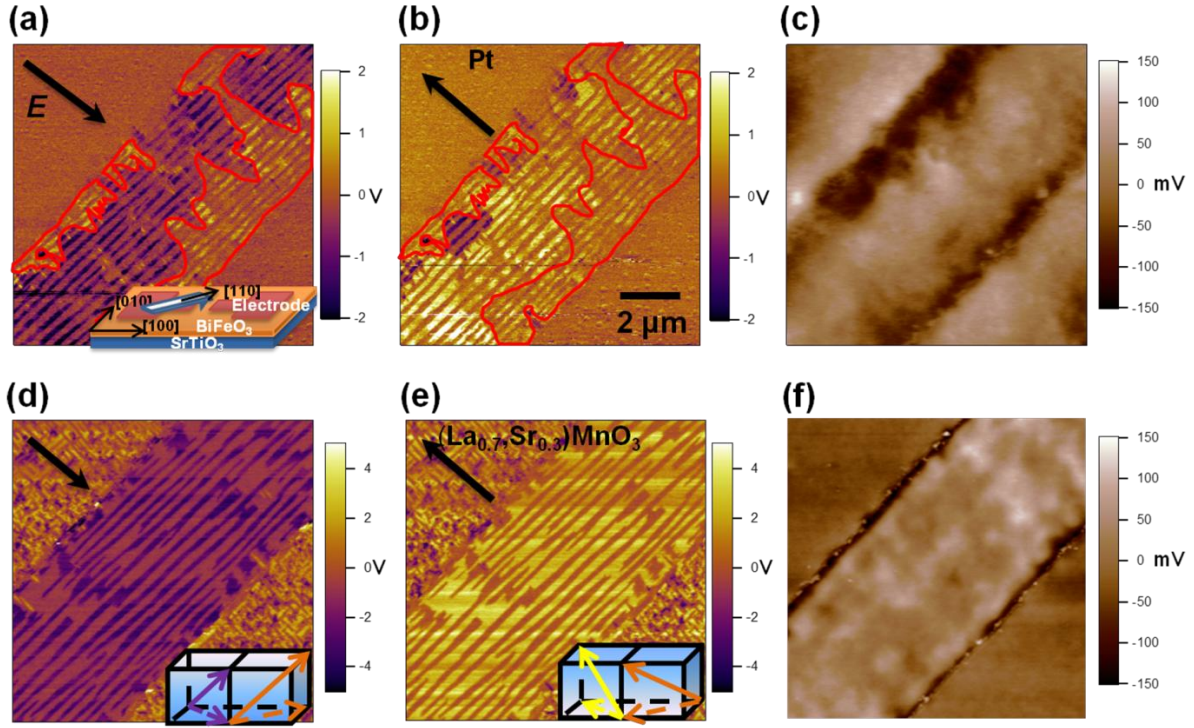

Figure 1, In plane (IP) PFM and SKPM images of planar Pt/BiFeO$_3$/Pt (a)-(c) and (La$_{0.7}$,Sr$_{0.3}$)MnO$_3$/BiFeO$_3$/(La$_{0.7}$,Sr$_{0.3}$)MnO$_3$ (d)-(f) devices after $10^{10}$ cycles of switching. (a), (b) Domain pining (outlined) across the whole Pt/BiFeO$_3$/Pt channel can be observed in the IP-PFM images. (c) Considerable charge injection (dark region) is observed at the Pt/BiFeO$_3$ interfaces from the SKPM image. (d)-(f) Negligible domain pining and charge injection are observed when (La$_{0.7}$,Sr$_{0.3}$)MnO$_3$ electrodes are used. Inset of (a) shows the planar device structure. The insets of (d) and (f) indicate the color code of IP-PFM images.

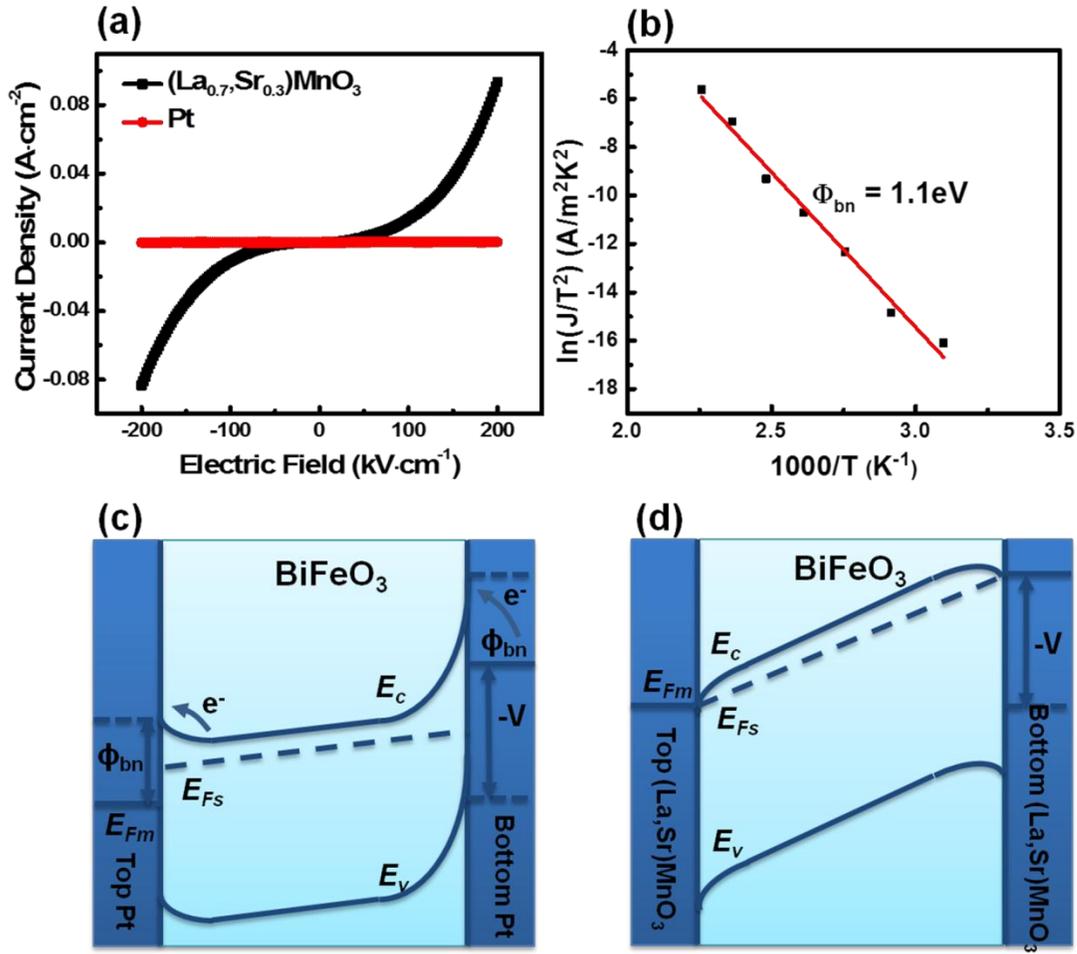

Figure 2, Leakage current density and energy band diagrams of planar BiFeO₃ devices. (a) The *I-V* curves show higher leakage current density when (La$_{0.7}$,Sr$_{0.3}$)MnO₃ electrodes are used, indicating a lower energy barrier at the (La$_{0.7}$,Sr$_{0.3}$)MnO₃/BiFeO₃ interface. (b) Fitting of *I-T* curve leads to a barrier height of ~1.1 eV for the Pt/BiFeO₃/Pt device. (c) A large Schottky barrier is expected at the Pt/BiFeO₃ interface. A large voltage drop arises at the reversely biased interface under external electric field, which drives electrons into the BiFeO₃ film. (d) A flat band is expected at the (La$_{0.7}$,Sr$_{0.3}$)MnO₃/BiFeO₃ interface and external field drops uniformly across the film.

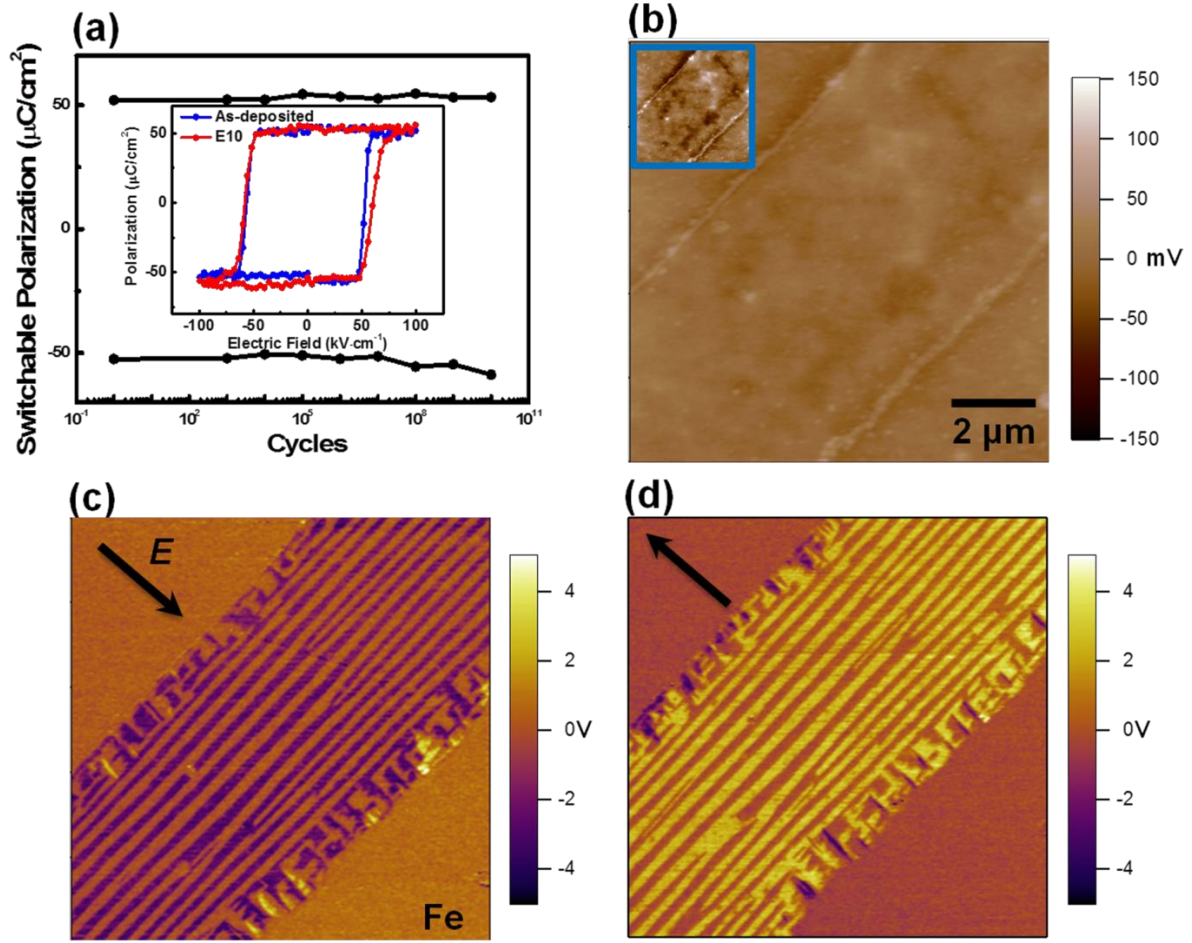

Figure 3, Fatigue performance of planar Fe/BiFeO₃/Fe device. (a) Switchable polarization vs. cycling number indicates no macroscopic fatigue up to $10^{10}$ cycles. Inset is the remanent hysteresis loops before and after electrical switching. SKPM image indicates negligible charge injection (b) and IP-PFM images show a very small amount of domain pinning (c)-(d) at the Fe/BiFeO₃ interface after $10^{10}$ switching cycles. Inset of (b) shows the same SKPM image with reduced scale.

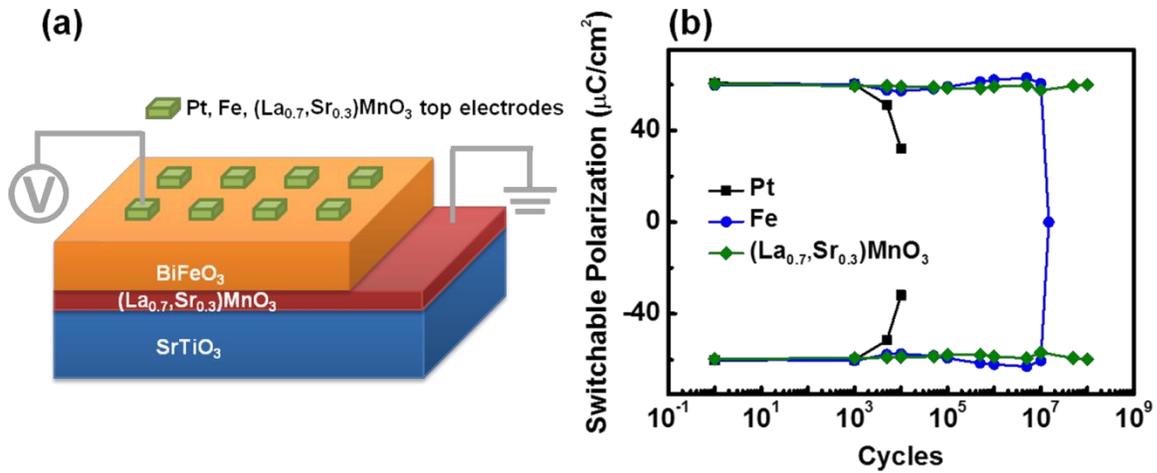

Figure 4, Fatigue performance of vertical BiFeO$_3$ capacitors. (a) Schematic illustration of the device structure. (b) Comparison of fatigue behaviors using three different top electrodes. Fatigue appears after $10^4$ cycles when Pt is used as top electrode, whereas both Fe and (La$_{0.7}$,Sr$_{0.3}$)MnO$_3$ top electrodes significantly improve the BiFeO$_3$ fatigue performance.